\documentstyle[preprint,aps,prb]{revtex}

\begin{document}
\title{Analysis of the approximations used in the van Hove-BCS theory.}
\author{D. Quesada$^{{\rm *}}$, A. Rubio-Ponce, and R. Baquero}
\address{Departamento de F\'{\i }sica,CINVESTAV, A. Postal 14-740, 07000 M\'{e}xico,\\
D.F.}
\author{R. Pe\~{n}a and C. Trallero-Giner}
\address{Departamento de F\'{\i }sica Te\'{o}rica, Universidad de La Habana, 10400 La%
\\
Habana, Cuba}
\maketitle

\begin{abstract}
An analysis of the different approaches used within the van Hove BCS model
for the high temperature superconductors has been done. How far the
employment of an asymptotic expression for the density of states
underestimates the thermodynamic parameters, as for example, the critical
temperature, the gap at zero temperature, and the known universal ratios of
the BCS model, is discussed. Analytical expressions obtained for some
thermodynamic functions are compared with the numerical results for which
the exact 2D density of states has been considered. Approximate analytical
expressions for the temperature dependence of the reduced gap, the
superconducting electronic specific heat, and the critical magnetic field
have been obtained and compared  with the corresponding numerical results.
The validity of the different approximations and of the equations obtained
for the thermodynamic functions in the framework of the van Hove model are
also analyzed.


\end{abstract}

\section{Introduction}

\bibliographystyle{unsrt}
\bibliography{acompat,example}

\begin{references}
\bibitem{eliashberg}  G.M. Eliashberg, J. Exp. Theoret. Phys. {\bf 11}, 966
(1960).

\bibitem{carbote}  J.P. Carbotte, Rev. Mod. Phys. {\bf 62}, 1067 (1990).

\bibitem{baquero}  J.M. Daams, J.P. Carbotte, M. Ashraf, and R. Baquero, J.
Low. Temp. Phys. {\bf 55}, 1 (1984); R. Baquero and J.P. Carbotte, J. Low.
Temp. Phys. {\bf 51}, 135 (1983).

\bibitem{mechan}  J. Ruvalds, Supercond. Sci. Technol. {\bf 9}, 905 (1996);
A.S. Davidov, Phys. Rep. {\bf 190}, 191 (1990).

\bibitem{friedel}  J. Friedel, J. Phys. (Paris) {\bf 48}, 1787 (1987); J.
Phys. Condens. Matter {\bf 1}, 7757 (1989).

\bibitem{labbe}  J. Labbe and J. Bok, Europhys. Lett. {\bf 3}, 1225 (1987);
J. Bok, Physica C {\bf 209}, 107 (1993).

\bibitem{tsuei1}  C.C. Tsuei, D.M. Newns, C.C. Chi, and P.C. Pattnaik, Phys.
Rev. Lett. {\bf 65}, 2724 (1990); D.M. Newns, C.C. Tsuei, P.C. Pattnaik, and
C.L. Kane, Comments Cond. Mat. Phys. {\bf 15}, 273 (1992).

\bibitem{mark}  R.S. Markiewicz, J. Phys. Condens. Matter {\bf 2}, 665
(1990); Physica C {\bf 168}, 195 (1990).

\bibitem{get}  J.M. Getino, M. de Llano, and H. Rubio, Phys. Rev. B {\bf 48}%
, 597 (1993).

\bibitem{goi}  A.G. Goicochea, Phys. Rev. B {\bf 49}, 6864 (1994).

\bibitem{davidq}  R. Baquero, D. Quesada, and C. Trallero-Giner, Physica C 
{\bf 271}, 122 (1996).

\bibitem{grassme}  R. Grassme and P. Seidel, J. Supercond. {\bf 9}, 619
(1996).

\bibitem{tsuei3}  D.M. Newns, C.C. Tsuei, and P.C. Pattnaik, Phys. Rev. B 
{\bf 52}, 13611 (1995).

\bibitem{liu}  M. Liu, D.Y. Xing, and Z.D. Wang, Phys. Rev. B {\bf 55}, 3181
(1997).

\bibitem{cucolo}  A.M. Cucolo, C. Noce, and A. Romano, Phys. Rev. B {\bf 53}%
, 6764 (1996).

\bibitem{band}   W.E. Pickett, Rev. Mod. Phys. {\bf 61}, 433 (1989); Z.X.
Shen and D.S. Dessau, Phys. Rep. {\bf 253}, 1 (1995).

\bibitem{roger}  E.N. Economou, $Green$ $functions$ $in$ $quantum$ $physics$%
, Springer Verlag, Berlin (1983).

\bibitem{agramo}  M. Abramowits and I. Stegun, Handbook of Mathematical
Functions, Dover Publications Inc. N.Y. (1970).

\bibitem{hira}  T. Hirata and Y. Asada, J. Supercond. {\bf 4}, 2 (1991).

\bibitem{muhlsch}  B. M\"{u}hlschlegel, Z. Phys. {\bf 155}, 313 (1959).

\bibitem{rizhyk}  I.S. Gradsteyn and I.M. Ryzhik, Table of Integrals, Series
and Products, Academic Press, San Diego (1980).

\bibitem{junod}  A. Junod in: $Physical$ $properties$ $of$ $high$ $%
temperature$ $superconductors$ $II$, ed. by D.M. Ginsberg, p. 13, World
Scientific, Singapore (1990).

\bibitem{schri}  J.R. Schrieffer, $Theory$ $of$ $superconductivity$,
W.A.Benjamin Inc. (1964).

\bibitem{krezin}  B.T. Geilikman, V.Z. Kresin, and N.F. Masharov, J. Low.
Temp. Phys. {\bf 18}, 241 (1975).

\bibitem{momono}  N. Momono and M. Ido, Physica C {\bf 264}, 311 (1996); N.
Momono, M. Ido, T. Nakano, and M. Ota, Physica C {\bf 235-240}, 1739 (1994).

\bibitem{mason}  T.E. Mason, G. Aeppli, S.M. Hayden, A.P. Ramirez, and H.A.
Mook, Phys. Rev. Lett. {\bf 71}, 919 (1993).

\bibitem{ausloos}  S. Dorbolo, M. Houssa, and M. Ausloos, Physica C {\bf 267}%
, 24 (1996).

\bibitem{hao}  J.H. Cho, Z. Hao, and D.C. Johnston, Phys. Rev. B {\bf 46},
8679 (1992).
\end{references}

For electron-phonon (e-ph) superconductors, the Eliashberg equations (EE) 
\cite{eliashberg} allow the exact calculation of the temperature dependence
of the thermodynamic functions. To solve EE one needs to know the e-ph
spectral distribution function, $\alpha ^{2}F(\omega )$, and the coulomb
repulsion pseudopotential parameter, $\mu ^{\ast }$. The function $\alpha
^{2}F(\omega )$ has been obtained from the inversion of tunneling data and
calculated theoretically for several e-ph superconductors. As a rule, a very
good agreement has been obtained between experimental and theoretical
calculations. The parameter $\mu ^{\ast }$ is more difficult to deal with in
a precise way. The common practice is to fit it to the experimental critical
temperature, $T_{c}$, through the linearized EE, valid at $T_{c}$. These
data are then used to solve the non-linearized EE from which the free-energy
difference as a function of the temperature, $T$, follows \cite
{carbote,baquero}.

BCS theory is the first solution found that gave the clue to the explanation
of the e-ph superconductivity. It is a weak coupling limit that, as it is to
be expected, deviates sometimes strongly, from the experimental results even
for some medium e-ph coupling superconductors \cite{carbote,baquero}. An
important role, as reference values, has been played by the universal ratios
(UR), $R_{1}=2\Delta (0)/k_{B}T_{c}$, $R_{2}=\Delta C(T_{c})/C_{en}(T_{c})$,
and $R_{3}=H_{c}(0)/\sqrt{N(0)}\Delta (0)$, where $\Delta (T)$ is the gap
function, $\Delta C(T_{c})=C_{es}(T_{c})-C_{en}(T_{c})$ is the jump of the
electronic specific heat at $T_{c}$, given by the difference between its
value in the superconducting state $C_{es}(T_{c})$ and in the normal state $%
C_{en}(T_{c})$, $H_{c}$ is the thermodynamic critical magnetic field, and $%
N(0)$ is the density of states (DOS) at the Fermi energy. For e-ph
conventional superconductors, the BCS UR do not depend on any parameter: $%
R_{1}=3.52$, $R_{2}=1.43$, and $R_{3}=2\sqrt{\pi }$. The  BCS theory was
mainly used as a reference, as a limit that helped characterize the relative
strength of the e-ph coupling for conventional superconductors (CS).

In the new high-Tc superconductors (HTS), the situation is quiet different
since the mechanism driving the superconducting phase transition is still
unknown. Actually, it is amazing how much it has been established about HTS
without knowing the mechanism. Two main attitudes have been adopted\cite
{mechan}: to assume a mechanism and to calculate the thermodynamics that
follows, or to incorporate some known facts into BCS theory which makes no
assumption on the specific mechanism. Several known facts
have been discussed within the BCS framework as, for example, the symmetry
of the gap and the influence of the dimensionality on the phenomenon of
superconductivity. In this sense, the van Hove BCS (v-BCS) scenario has been
formulated by different authors \cite{friedel,liu}. Both, analytical and
numerical treatments have been carried out leading to results that present
noticeable differences in the thermodynamic functions. Tunneling experiments
and the dependence of the specific heat on temperature \cite{cucolo}, have
been analyzed following the BCS formalism. For example, the contributions
that come from the CuO$_{\text{2}}$ planes, CuO chains, and c-axis, have
been decoupled\cite{cucolo}. The one coming from the CuO$_{\text{2}}$ plane
seems to be the most important one. Most of the work has been devoted to it.
In this sense, it is necessary to understand very well superconductivity on
the CuO$_{\text{2}}$ plane and to analyze in detail how the different
approximations used describe the phenomenon. In other words, it is important
to be aware whether certain deviations from the experimental results come
from physics or from mathematics.

In this paper we want to analyze three types of solutions to the van Hove
scenario, {\em i.e.}, the analytic ones and their approximations, the
numerical ``asymptotic'' solution that makes use of the asymptotic behavior
of the two-dimensional density of states, and what we will call the
``exact'' solution where the full elliptic integral of the first kind,
taking place in the 2D DOS, is considered and treated numerically. The
calculations presented here assume an s-wave symmetry for the gap in the CuO$%
_{\text{2}}$ plane since it is the most widely studied case, although it is
known not to be the experimental verified result. Nevertheless, this is not
important for the sake of our purpose and the analysis obtained does not
change in any essential way whether we introduced or not a d-wave or a
mixing of s and d gap symmetries.

The rest of the paper is organized as follows. In sec. 2, we compare the
analytic, the asymptotic and the exact solution within the van Hove scenario
for the gap, the critical temperature, the temperature dependence of the
gap, the electronic specific heat, and the thermodynamic critical field.
Discussions and conclusions are considered in sec. 3.

\section{BCS model with a van Hove singularity in the DOS}

The layered structure of the new superconducting materials leads to almost
dispersionless electronic spectrum on the c-axis \cite{band} and therefore,
as a first approximation, it can be modeled as a stack of decoupled CuO$_{2}$
planes. This constitutes the so called van Hove scenario. For the CuO$_{2}$
planes, we take the following electronic dispersion relation: 
\begin{equation}
\varepsilon _{\vec{k}}=-2t(\cos k_{x}a+\cos k_{y}a)+4t^{\prime }\cos
(k_{x}a)\cos (k_{y}a)\text{ \ },
\end{equation}
where $t$ and $t^{\prime }$ are the coupling parameters between the Cu-O
atoms and O-O atoms respectively, $k_{x}$, $k_{y}$ are wave vector
components, and $a$ is the lattice constant. The origin of the energy is at
the Fermi level, $\varepsilon _{F}$. The resulting topology of the
electronic band structure in the first Brillouin zone for $t^{\prime }=0$
(a) and $t^{\prime }=0.4t$ (b) are shown in Fig. 1. The saddle points at $%
\vec{k}=(\pi /a,0)$ and $(0,\pi /a)$ are clearly seen yielding to the van
Hove singularity (vHs) which enhances the DOS. The parameter $t^{\prime }$
ranges between 0 and 0.5 $t$ and it is responsible for the observed changes
in the shape of the Fermi surface (compare Fig. 1(a) to Fig. 1(b)). It is
worth noticing that for a fixed carrier concentration, the parameter $%
t^{\prime }$ drives the relative position of the singularity with respect to 
$\varepsilon _{F}$. Later on, we will show how almost all superconducting
properties depend on this relative position.

The single spin DOS for a stack of layers \cite{roger} is equal to 
\begin{equation}
N(\varepsilon )=\frac{S \ n_p}{a^{2}}\frac{N_{0}}{\sqrt{1+\frac{%
\varepsilon \ t^{\prime }}{t^{2}}}}\ K\left[ \sqrt{\frac{16t^{2}-(%
\varepsilon -4t^{\prime })^{2}}{16t^{2}(1+\frac{\varepsilon t^{\prime }}{%
t^{2}})}}\right] \text{ \ },  \label{edos}
\end{equation}
where $K(m)$ is the elliptic integral of the first kind \cite{agramo}. S is
the area of the sample, $n_{p}$ is the number of layers, and $N_{0}=1/2\pi
^{2}t$. The asymptotic form for $K(m)$ in the neighborhood of the
singularity with $t^{\prime }=0$ is \cite{agramo} 
\begin{equation}
N(\varepsilon )\approx \frac{Sn_{p}}{a^{2}}\ \frac{N_{0}}{2}\ \ln \left( 
\frac{16}{|\varepsilon /4t|}\right) \text{ \ }.  \label{ados}
\end{equation}
The equation above allows us to obtain analytical expressions for some of
the thermodynamic functions. The BCS theory equation for the gap is 
\[
\Delta _{\vec{k}}(T)=\frac{1}{2}\ \sum_{\vec{k}^{\prime }}\ V_{\vec{k},\vec{k%
}^{\prime }}\ \frac{\Delta _{\vec{k}^{\prime }}}{E_{\vec{k}^{\prime }}}\
\tanh \left( \frac{E_{\vec{k}^{\prime }}}{2k_{B}T}\right) \text{ \ }, 
\]
where $E_{\vec{k}^{\prime }}=\sqrt{\varepsilon _{\vec{k}^{\prime
}}^{2}+|\Delta _{\vec{k}^{\prime }}(T)|^{2}}$, and $V_{\vec{k},\vec{k}%
^{\prime }}$ is the effective electron-electron interaction matrix element.
In agreement with the BCS formulation we use the following parametrization
for $V_{\vec{k},\vec{k}^{\prime }}$ 
\[
V_{\vec{k},\vec{k}^{\prime }}=\left\{ 
\begin{array}{ll}
V_{1}\ ;\  & \text{ \thinspace }%
\mbox{ if $|\varepsilon_{\vec k}|,\ |\varepsilon_{\vec k'}| <
\varepsilon_c$} \\ 
0\;\;\,;\  & \;\;\mbox{otherwise}\hspace{2cm},
\end{array}
\right. 
\]
and $\varepsilon _{c}$ is an energy cut-off. Following the standard BCS
procedure, we get 
\begin{equation}
\frac{2}{V_{1}}=%
\displaystyle\int %
\limits_{-\varepsilon _{c}}^{\varepsilon _{c}}d\varepsilon \text{ }%
N(\varepsilon )\text{ }\frac{\tanh \left( \sqrt{\varepsilon ^{2}+\Delta
^{2}(T)}/2k_{B}T\right) }{\sqrt{\varepsilon ^{2}+\Delta ^{2}(T)}}\text{ \ }.
\label{eqgap}
\end{equation}

\subsection{The gap and the critical temperature}

The numerical solutions of eq. (\ref{eqgap}) with a DOS given by eq. (\ref
{edos}) will be referred to as the ``exact'' solution. When eq. (\ref{ados})
is used instead, we will talk about ``asymptotic solution''. We, first,
compare the exact to the asymptotic DOS in the calculation of $T_{c}$ and $%
\Delta (0)$, in particular, their dependence on the dimensionless
interaction parameter $\lambda \equiv N_{0}V$, with $V=V_{1}n_{p}S/a^{2}\ $.
We present these results in Fig. 2. Here, $\alpha \equiv 2t^{\prime }/t$,
twice the O-O second nearest neighbors hopping parameter in units of the
Cu-O first nearest neighbors one. In Fig. 2(a), we present the dependence on 
$\lambda $ of the critical temperature $T_{c}$ and in Fig. 2(b), the one of
the gap $\Delta (0)$, normalized to the cut-off energy $\varepsilon _{c}$.
The curves correspond to $\alpha =0.01$, $0.04$, and $0.07.$ Both
parameters, $T_{c}$ and $\Delta (0)$, increase with $\lambda $ almost in the
same way. For lower values of $\alpha $ (Cu-O interaction stronger than the
O-O one), $T_{c}$ and $\Delta (0)$ both are a very steep function of $%
\lambda $. Hence, for a single plane, the dependence of \thinspace the
critical temperature and the gap on $\lambda $ is strong. It is worth
noticing that $T_{c}$ of order 100K can be reached with $\lambda =0.15$ and $%
\alpha =0.01$. On the other hand, for higher values of $\alpha $ (a big
relative hopping energy to the O-O atoms), $i.e$ $\alpha =0.07$, $\lambda $
would have to increase as high as 0.25 to obtain $T_{c}\approx 100$K. It is
worth noticing that for $\alpha =0.07$, $\lambda $ needs to be of the order
of 0.30 to get $\Delta (0)/\varepsilon _{c}\approx 1$. 

In Fig. 3, we show the exact solution for the universal ratio, $%
R_{1}\equiv 2\Delta (0)/k_{B}T_{c}$ as a function of $\lambda $, for the
same values of $\alpha $ considered above. At $\lambda $=0.15 and $\alpha $%
=0.01, $R_{1}$=3.82; while for higher values of $\lambda $, $R_{1}$ tends to
4 irrespective of the parameter $\alpha $. Notice that, according to these
results, the asymptotic value is reached for $\lambda $'s that might be very
high compared to the expected ones in the cooper oxide superconductors \cite
{band}. On the other hand, at low $\lambda $ values, we get the usual BCS
value, {\it i.e.}, $R_{1}=3.52$. On the overall, these results show that for
any value of $\lambda $ and any chosen parameter $\alpha $, $3.5\leq
R_{1}\leq 4$. Therefore, in any case, the 2D character of superconductivity
in the cooper oxides rises $R_{1}$ by at most 0.5 within the BCS formulation.

Approximate analytic solutions are obtained by inserting eq. (\ref{ados})
into eq. (\ref{eqgap})\cite{davidq}: 
\begin{equation}
\Delta (0)=64t\ \exp \left( 1-\sqrt{\frac{4}{N_{0}V}+\ln ^{2}\left( \frac{%
\varepsilon _{c}}{64t}\right) -1}\right) \text{ \ },
\end{equation}
\begin{equation}
k_{B}T_{c}=32t\ \exp \left( 1-\sqrt{\frac{4}{N_{0}V}+\ln ^{2}\left( \frac{%
\varepsilon _{c}}{64t}\right) -1}\right) \text{ \ },
\end{equation}
from which a parameter--independent universal ratio follows: 
\begin{equation}
R_{1}=\frac{2\Delta (0)}{k_{B}T_{c}}=4.
\end{equation}
The numerical calculations using the {\it exact} DOS converge to $R_{1}=4$
for high values of $\lambda $, so the analytic solution (7) , represents an
upper bound for this ratio (see Fig. 3). The $R_{1}$ value is smaller than
those reported experimentally, where $R_{1}$ ranging between 4 and 8 are
found \cite{hira}. This disagreement is well outside the experimental
error.

In Fig. 4, the dependence of $T_{c}$ on the dimensionless interaction
parameter, $\lambda $, is studied. The solid lines are numerical solutions
obtained from eq. (\ref{eqgap}) with the DOS given by eq. (2) and $t^{\prime
}=0$. The dot--dashed lines correspond to solutions using the asymptotic
behavior of the DOS (eq. (3)). Three set of curves are presented in each
case: $\varepsilon _{c}=$ 20 meV (1), 35 meV (2), and 50 meV (3). As it can
be seen, the cut-off energy has a strong influence on $T_{c}$. At $\lambda =$
0.20, for example, the solid lines range between $100K$ and $300K$ for $%
20 meV \leq \varepsilon _{c}\leq 50 meV$. It is difficult to decide for a
precise value of $\varepsilon _{c}$ which actually acts as a free parameter
and quantitative results are subject to a physical justification. In Fig. 4,
we also show the results using the {\it exact} DOS (solid lines). Compare
them to the asymptotic expression (dot--dashed lines). The {\it exact} DOS
gives, for the same value of $\varepsilon _{c}$, always higher values for $%
T_{c}$. At $\lambda $=0.15, for example, the {\it exact} $T_{c}$ is about
twice the one obtained from the asymptotic DOS given by eq. (\ref{ados}).
Furthermore, the analytic solution given by eq. (6) using $\varepsilon
_{c}=50$meV (dashed line) gives values for T$_{c}$ that are just (10\%)
above the numerical results using the asymptotic DOS but quite lower than
the exact solution. So, the effect of the vHs is quite underestimated by
using in either way the asymptotic form for $K(m)$. Any analysis of HTS
based on BCS theory should take into account these very strong dependences
on several parameters before reaching any even qualitative conclusion. This
is a main result of this paper. We will illustrate it further.

\subsection{Temperature dependence of $\Delta(T)/\Delta(0)$}

The temperature dependence of the reduced gap, $\Delta (T)/\Delta (0)$, on
the reduced temperature $T/T_{c}$ is presented in Fig. 5. For the sake of
comparison the M\"{u}hlschlegel model\cite{muhlsch} is also shown(in
circles). It is found that the temperature dependence of $\Delta (T)/\Delta
(0)$ in the framework of the models considered here present the known
universal character irrespective of the value of the parameters $\alpha $, $%
\varepsilon _{c}$ and $\lambda $.

In the following, approximate analytic expressions for the gap at low
temperatures and close to $T_{c}$ are derived. In the low temperature regime
and taking $t^{\prime }=0$, eq. (\ref{eqgap}) can be cast in the form $%
I_{1}-I_{2}=-2I_{3}$, with 
\[
I_{1}=\int\limits_{-\varepsilon _{c}}^{\ \varepsilon _{c}}\!d\varepsilon
N(\varepsilon )\frac{1}{\sqrt{\varepsilon ^{2}+\Delta ^{2}(0)}}\
;\;\;\;\;I_{j}=%
\displaystyle\int %
\limits_{-\varepsilon _{c}}^{\ \varepsilon _{c}}\!d\varepsilon N(\varepsilon
)\frac{[f(E)]^{j-2}}{\sqrt{\varepsilon ^{2}+\Delta ^{2}(T)}}\ \ \,(j=2,3)%
\text{ \ }, 
\]
where $f(E)$ is the Fermi-Dirac distribution function. Using eq. (\ref{ados}%
), the difference $I_{1}-I_{2}$ is calculated in a similar way to that used
in ref. \cite{davidq}:

\[
I_1-I_2\propto (1+\ln 16)\ \ln \left( \frac{\Delta (T)}{\Delta (0)}\right) +%
\frac 12\ln ^2\left( \frac{\Delta (0)}{4t}\right) -\frac 12\ln ^2\left( 
\frac{\Delta (T)}{4t}\right) \text{ \ }, 
\]
\begin{equation}
I_3\propto \left[ \ln (16)\sqrt{\frac{\pi k_BT}{2\Delta (0)}}-\ln \left( 
\frac{\Delta (0)}{4t}\right) +1\right] \exp \left( -\frac{\Delta (0)}{k_BT}%
\right) \text{ \ }.  \nonumber
\end{equation}
Collecting the results for $I_1-I_2$ and $I_3$ and taking into account that $%
R_1=4$, the following analytical expression for the gap, $\Delta (T)$, is
obtained 
\begin{equation}
\frac{\Delta (T)}{\Delta (0)}=1-\frac 34\left( \sqrt{\frac{\pi T}{T_c}}+%
\frac{10}3\right) \exp \bigg(-\frac{2T_c}T\bigg)\text{ \ }.  \label{gap0}
\end{equation}
Let us look now at the behavior of the gap near $T_c$. For this purpose eq. (%
\ref{eqgap}) is rewritten as $K_1-K_2=K_3$ with

\[
K_1=\int\limits_0^{\ \omega }dx\ \ln \left( \frac{16}x\right) \ \frac{\tanh
(\beta _cx/2)}x\hspace{0.3cm}, 
\]

\[
K_2=\int\limits_0^{\ \omega }dx\ \ln \left( \frac{16}x\right) \ \frac{\tanh
(\beta x/2)}x\text{ \thinspace \thinspace }, 
\]
\begin{equation}
K_3=-4\beta ^3y^2\sum_{n=1}^\infty 
\displaystyle\int %
\limits_0^{\ \omega }dx\ \ln \left( \frac{16}x\right) \ \frac 1{[(2n-1)^2\pi
^2+\beta ^2x^2]^2}\text{ \ },  \nonumber
\end{equation}
where $\beta _c=4t/k_BT_c$, $x=\varepsilon /4t$, $y=\Delta (T)/4t$, $\omega
=\varepsilon _c/4t,$ $\beta =4t/k_BT$, and the following identity has been
used 
\[
\tanh \left( \frac b2\right) =4b\sum_{n=1}^\infty \frac 1{(2n-1)^2\pi ^2+b^2}%
\text{ \ .} 
\]
Since $K_3$ decreases strongly for large values of $\beta x$, the upper
limit can be extended to infinity and

\begin{equation}
K_3=\left( \frac{\Delta (T)}{\pi k_BT}\right) ^2\sum_{n=1}^\infty \left[ 
\frac{\ln (2n-1)}{(2n-1)^3}-\frac{\ln (16\beta /\pi )+1}{(2n-1)^3}\right] 
\text{ \ }.  \label{k3}
\end{equation}
The difference $K_1-K_2$ can be obtained using the same approach of ref. 
\cite{davidq} and results in

\begin{equation}
K_1-K_2=(1+\ln 16)\ \ln \left( \frac T{T_c}\right) -\frac 12\left[ \ln
^2\left( \frac{k_BT}{2t}\right) -\ln ^2\left( \frac{k_BT_c}{2t}\right) %
\right] \text{ \ }.  \label{k12}
\end{equation}
The above equations are the starting point to develop an explicit equation
for the gap close to $T_c$. Differentiating $K_1-K_2$ in eq. (\ref{k12}) and 
$K_3$ from eq. (\ref{k3}) with respect to $T$ results in 
\begin{equation}
\frac{d\Delta ^2(T)}{dT}\Bigg|_{T_c}=-\frac{8\pi ^2}{7\zeta (3)}%
k_B^2T_c=-9.38k_B^2T_c  \label{dDT}
\end{equation}
which is the same as c-BCS result. To derive the above equation the
following identities have been employed \cite{rizhyk}

\[
\sum_{n=1}^{\infty }\frac{1}{(2n-1)^{k}}\;=(1-2^{-k})\zeta (k)\text{ \ .} 
\]
\begin{equation}
\sum_{n=1}^{\infty }\frac{\ln (2n-1)}{(2n-1)^{k}}=-2^{-k}\zeta (k)\ \ln
2-(1-2^{-k})\zeta ^{\prime }(k)\text{ \ .}  \nonumber
\end{equation}
where $\zeta (k)$ is the Riemann Zeta function, and $\zeta ^{\prime }(k)$
its derivative. From eqs. (\ref{k3}) and (\ref{k12}) follow the functional
dependence $\Delta ^{2}(T)=F(T)$. Irrespective to the form that the function 
$F(T)$ has, for $T$ near $T_{c}$ and due to the smallness of $\Delta ^{2}(T)$
, this can be expand in powers of $(T-T_{c})$, thus 
\[
\Delta ^{2}(T)=\left( \frac{dF}{dT}\right) \Bigg|_{T_{c}}(T-T_{c})=\left( 
\frac{d\Delta ^{2}(T)}{dT}\right) \Bigg|_{T_{c}}(T-T_{c})\text{ \ }. 
\]
From eq. (\ref{dDT}), $\Delta ^{2}(T)$\thinspace \thinspace can be cast into 
\[
\Delta ^{2}(T)=9.38\ k_{B}^{2}T_{c}^{\text{ }2}\ \left( 1-\frac{T}{T_{c}}%
\right) 
\]
and taking into account eq. (7) it is obtained that 
\begin{equation}
\frac{\Delta (T)}{\Delta (0)}=1.53\ \sqrt{1-\frac{T}{T_{c}}}\text{ \ }.
\label{DTD0}
\end{equation}
The expressions for $\Delta (T)$ given by eqs. (\ref{gap0}) and (\ref{DTD0}%
), are plotted in Fig. 5 in dashed and dot-dashed lines, respectively. It
can be seen that eq. (\ref{gap0}) is valid in the range $0\leq T/T_{c}\leq
0.4$, whereas eq. (\ref{DTD0}) in the range $0.95\leq T/T_{c}\leq 1$.

\subsection{Electronic specific heat}

The fermion specific heat, $C_{en}$, in the normal state is given by 
\begin{equation}
C_{en}(T)=\frac{2}{k_{B}T\text{ }^{2}}\int\limits_{-4t-4t^{\prime }}^{\
4t-4t^{\prime }}d\varepsilon \text{ }N(\varepsilon )\text{ }\varepsilon ^{%
\text{ }2}f(\varepsilon )\text{ }[1-f(\varepsilon )]\text{ \ }.  \label{cen}
\end{equation}
In Fig. 6(a) the $C_{en}(T)/T$ dependence on $T$\ is presented for different
values of the parameter $\alpha $ and $\varepsilon _{F}=0$. For $\alpha =0$, 
$C_{en}/T$\ has a singular behavior as the temperature decreases. On the
other hand, for $\alpha >$ $0.03$ and very low temperatures, the electronic
specific heat presents a linear behavior with $T$. These features are
consistent with those reported in ref. \cite{junod}, where the usual
constant coefficient in the linear term of the specific heat becomes a function
of temperature.

An approximate analytic expression for $C_{en}$ will be now derived. After
inserting eq. (\ref{ados}) in eq. (\ref{cen}) we get 
\begin{equation}
C_{en}(T)=\frac{\pi ^2}3RN_0\ k_BT\left[ 1.73-\ln \left( \frac{k_BT}{4t}%
\right) \right] \text{ \ ,}  \label{Cen}
\end{equation}
where $R$ is the universal gas constant, $n_pS/a^2$ is assumed to be equal
to the Avogadro's number, and the specific heat is calculated per mole. The
first term in eq. (\ref{Cen}) is the usual contribution from constant DOS.
The second one, $T\ln T$, is due to the vHs. A similar dependence has been
reported in. ref.\cite{goi}.

The superconducting-state electronic specific heat can be written as 
\begin{equation}
C_{es}(T)=\frac 2{k_BT^2}\int\limits_{-\varepsilon _c}^{\ \varepsilon
_c}d\varepsilon \ N(\varepsilon )f(E)[1-f(E)][E^2-\frac 12T\frac{d\Delta
^2(T)}{dT}]\text{ \ }.  \label{Ces}
\end{equation}
For the derivative of the square of the gap as a function of $T$ we use eq. (%
\ref{dDT}). We have calculated the dependence of $C_{es}(T)/C_{en}(T_c)$ on $%
T/T_c$, for three different values of $\alpha $, ($\varepsilon _c=20$ meV, $%
\lambda =0.1$). We get a universal behavior consistent with the one obtained
for $\Delta (T)/\Delta (0)$.

Inserting the eq. (\ref{ados}) into eq. (\ref{Ces}) results in 
\[
C_{es}(T)=4RN_{0}\ k_{B}T_{c}\ \sqrt{2\pi \left( \frac{2T_{c}}{T}\right) ^{3}%
}\ \exp \left( -\frac{2T_{c}}{T}\right) \left[ 3.46+\frac{1}{4}\ln \left( 
\frac{2T_{c}}{T}\right) \right] \text{ \ },
\]
where a value of $k_{B}T_{c}/4t=0.015$ has been assumed. This approximated
analytical equation is valid at low temperatures $(0<T/T_{c}<0.4)$ where the
gap is almost independent of temperature (see Fig. 5). Finally, for the
ratio $C_{es}(T)/C_{en}(T_{c})$  follows: 
\[
\frac{C_{es}(T)}{C_{en}(T_{c})}=7.15\sqrt{\left( \frac{2T_{c}}{\pi T}\right)
^{3}}\ \exp \left( -\frac{2T_{c}}{T}\right) \ \left[ 1.4+\frac{1}{10}\ \ln
\left( \frac{2T_{c}}{T}\right) \right] 
\]
Notice that the expected exponential behavior has been obtained.

The specific heat jump at $T_{c}$ is given by 
\begin{equation}
\Delta C(T_{c})=-\frac{1}{k_{B}T_{c}}\frac{d\Delta ^{2}(T)}{dT}\Bigg|%
_{T_{c}}\int\limits_{-4t-4t^{\prime }}^{\ 4t-4t^{\prime }}d\varepsilon \
N(\varepsilon )f(\varepsilon )[1-f(\varepsilon )]\text{ \ }.  \label{DTDc}
\end{equation}
In Fig. 6(b) we show the dependence of $R_{2}=\Delta C(T_{c})/C_{en}(T_{c})$
on $\alpha $. For low values of $\alpha $, $R_{2}$ approaches the asymptotic
behavior of 1.95 as the cut-off energy increases. At intermediate values of $%
\alpha $, R$_{1}$ crosses the c-BCS value towards lower values. At $\alpha =%
\frac{\varepsilon _{c}}{2t}$ we get a discontinuity (in disagreement with
ref. \cite{goi}). $\alpha =\varepsilon _{c}/2t$ corresponds to the vHs
energy shift from the Fermi level ($\varepsilon _{F}=0$). The jump observed
in the figure is due to the vHs (see eq. (\ref{ados})). Finally the curve
approaches asymptotically the c-BCS value for higher value of $\alpha $. At
higher values of $\alpha $ a crossover to the c-BCS value of $1.43$ is
reached.

In order to estimate the maximal value that $R_2$ can reach, we take $\alpha
=0$. Then, from eqs. (\ref{ados}), (\ref{dDT}), (\ref{Cen}), and (\ref{DTDc}%
) we get

\[
\frac{\Delta C(T_{c})}{T_{c}}=9.38\ RN_{0}k_{B}\left[ 1.96-\frac{1}{2}\ln
\left( \frac{k_{B}T_{c}}{4t}\right) \right] 
\]
and 
\[
\frac{\Delta C(T_{c})}{C_{en}(T_{c})}=2.85\left[ \frac{1.96-\frac{1}{2}\ln
(k_{B}T_{c}/4t)}{1.73\,\,-\,\,\ln (k_{B}T_{c}/4t)}\right] \text{ \ }. 
\]
Assuming $k_{B}T_{c}/4t=0.015$, we get an upper limit for $R_{2}$: 
\[
R_{2}=\frac{\Delta C(T_{c})}{C_{en}(T_{c})}=1.95\text{ \ }. 
\]
This result is slightly higher than the conventional BCS model value. The
difference represents the influence of the vHs. Even thought the value of
1.95 follows the trend to higher values in the UR, this is not high enough
to be in agreement with those reported from experimental estimations ($%
2<R_{2}<4.6$)\cite{junod}.

\subsection{Critical Magnetic Field}

Finally, the thermodynamic critical magnetic field, $H_{c}(T)$, can be
evaluated from the thermodynamic relationship 
\begin{equation}
\frac{H_{c}^{2}(T)}{8\pi }=F_{n}(T)-F_{s}(T)\text{ \ },  \label{hc}
\end{equation}
where the $F_{n}(T)(F_{s}(T))$ is the free energy in the normal
(superconducting) state\cite{schri}. We use eq. (2) in eq. (20) and get the
reduced temperature dependence of the reduced thermodynamic critical
magnetic field $H_{c}(T)/H_{c}(0)$that appears in Fig. 7 for two different
values of $\varepsilon _{c}$ and $\lambda $ at $\alpha =0$ (see figure
caption for details). Curve 4 is the parabolic behavior known for
conventional superconductors. Notice the strong departure from the parabolic
behavior.

After some straightforward but slightly lengthy algebra, we arrive at the
following expression 
\[
\frac{H_{c}^{2}(0)}{N_{0}/\Omega \ \Delta ^{2}(0)}=2\pi \left[ \frac{4}{3}%
-\ln \left( \frac{\Delta (0)}{64t}\right) \right] \text{ \ }, 
\]
where $\Omega =a^{2}d$ and $d$ is the interplanar distance. From the above
equation it follows that $H_{c}(0)$ has a similar behavior as $\Delta (0)$
as a function of $\varepsilon _{c}$, $\lambda $ and $\alpha $. Further,
using the inequality $-\ln x>1-x$, for $x<1$ and $\Delta (0)/64t\ll 1$, a
lower limit for the UR $R_{3}=H_{c}(0)$ $/\sqrt{N_{0}/\Omega }\Delta (0)$ is
obtained 
\[
R_{3}=\frac{H_{c}(0)}{\sqrt{N_{0}/\Omega }\Delta (0)}>2.16\sqrt{\pi }\text{
\ }. 
\]
This result represents another UR, whose value is higher than the 3D BCS
model $(2\sqrt{\pi }),$fixing a minimum for this ratio .

The temperature dependence of $H_c(T)$ near $T_c$, can be obtained in the
following way. Near T$_c$, $\Delta (T)$ is a small and $F_s(T)$ can be
expanded in powers of $\Delta ^2(T)$. To second order, we get:

\begin{equation}
F_{s}(T)-F_{n}(T)=-a_{1}(T_{c})\Delta ^{4}(T)\text{ \ },
\end{equation}
\begin{equation}
a_{1}(T_{c})=\frac{1}{2}\beta ^{3}%
\mathop{\displaystyle\sum}%
_{n\text{ },\text{ }\vec{k}}\text{ }\frac{1}{\left[ (2n-1)^{2}\pi ^{2}+\beta
_{c}^{2}\varepsilon _{\vec{k}}^{2}\right] ^{2}}\text{ \ .}  \label{a1}
\end{equation}
The sum over $\vec{k}$ in eq. (\ref{a1}), can be transformed into an
integral over the energy, and using eq. (\ref{k3}), we get 
\[
\frac{H_{c}^{2}(T)}{8\pi }=0.29\left( \frac{N_{0}\Delta ^{2}(0)}{a^{2}d}%
\right) \left[ 2.84-\ln \left( \frac{k_{B}T_{c}}{4t}\right) \right] \left( 1-%
\frac{T}{T_{c}}\right) ^{2}\text{ \ }. 
\]
From this expression it follows:

\begin{equation}
\frac{H_{c}(T)}{H_{c}(0)}=0.83\ \left( 1-\frac{T}{T_{c}}\right) \text{ \ }.
\label{HcH0}
\end{equation}
Equation (\ref{HcH0}) has a slope of 0.83 which is smaller than the one in
c-BCS (1.74). The approximate $H_{c}(T)$ expression for $T$ near $T_{c},$ of
eq. (\ref{HcH0}) is valid in the range $0.75\leq T/T_{c}\leq 1$.

\section{Discussion and Conclusions}

In the present paper the thermodynamics of the BCS model within the vHs
scenario is carried out using both the exact DOS and its asymptotic behavior
(analytically and numerically). We were interested in the intrinsic error
that derives from the use of the asymptotic approximation to the van Hove
DOS as compared to the exact one. For that purpose we assumed an s-wave
symmetry for the gap in spite of the experimentally shown d or (d+s)-wave
symmetry. For our purpose this does not make any difference.  Our study
helps ping point to specific differences with experiment that come from the
mathematical approximations used rather than from physics. A first
conclusion is that the widely spread use of the logarithmic asymptotic
behavior near the singularity of the elliptic integral of the frst kind that
is obtained for the DOS in the van Hove scenario is a very disputable
approximation. It can lead to results that differ substantially from the
ones obtained when the exact DOS is used. So most of the analytical
expressions (obtained in this way) are to be used with great care, in
general, even to draw from them qualitative conclusions.

Furthermore, even using the exact DOS of the van Hove scenario, we still get
(within v-BCS) a picture that does not always agree with experiment.

Numerical computations of the gap at zero temperature $\Delta (0)$ and $T_{c}
$ as a function of the dimensionless interaction parameter,$\ \ \lambda ,$
were carried out for different values of $\ \alpha \equiv 2t^{\prime }/t,$ 
with $\varepsilon _{c},$ the cut-off energy, fixed ( see Fig. 2). T$_{c}$
and $\Delta (0)$ increase with $\lambda $, but as $\alpha $ is increased,
with $\lambda $ fixed, a strong decrease in both T$_{c}$ and $\Delta (0)$ is
observed. As $\alpha $ increases, the range of possible values of $\lambda $
that are consistent with the experimental results, increase towards those
often found in CS. An analysis of \ Fig. 2 shows that for $\lambda >0.15$
the ratios $k_{B}T_{c}/\varepsilon _{c}$ and $\Delta (0)/\varepsilon _{c}$
are higher than one. In CS, this fact is understood as a signature of
lifetime effects \cite{carbote}. Thus, for small values of $\alpha $ and $%
\lambda >0.15$, many body effects should play an important role . This is
consistent with the explanation given for CS \cite{krezin} to account for
some high values found for $R_{1}$. The universal ratio, $R_{1}=2\Delta
(0)/k_{B}T_{c}$, ranges between 3.5 and 4 for any values of $\alpha $ and $%
\lambda $, well below the experimental reported value. The highest values of 
$\Delta (0)$ and $T_{c}$ are obtained at $\alpha =0$. We have derived
analytic expressions in this limit to examine further this point. The
asymptotic approximation underestimate $\Delta (0)$ and $T_{c}$ as compared
to the exact result. It is interesting to notice further that there is but a
small difference using Eq. 3 whether the problem is solved analytically or
numerically (only by 10 $\%$ , see Fig. 4).

The temperature dependence of the reduced gap $\Delta (T)/\Delta (0)$ on the
reduced temperature $T/T_{c}$ presents the known universal behavior
irrespective of the value of $\alpha $, $\varepsilon _{c}$ and $\lambda $.
Analytic expressions can be obtained for the gap at $0\leq T\leq 0.4$ (eq.
(9)) as well as at $0.95\leq T/T_{c}\leq 1$ (eq. (15)). The presence of the
vHs leads to higher values of $\Delta (0)$ and $T_{c}$ than those reported
for the CS even for low values of the dimensionless interaction constant $%
\lambda $. Furthermore, the magnitudes $\Delta (0)$, $T_{c}$ and $R_{1}$ are
strongly affected by the departure of the vHs from the $\varepsilon _{F}$.
This is an obvious result since the number of states around $\varepsilon
_{F} $ (that contribute to the superconductor state) decreases as the
singularity moves far away from the Fermi level.

We have further calculated the specific heat. Our results appear in Figs.
6(a) and 6(b). We get for $C_{es}(T)/T$ an exponential decay at low
temperature (for $a$ $>0$) in contrast with that reported in \cite
{junod,momono,mason}. It is possible to show, that the corresponding width
\thinspace in the transition region ($T$ close to $T_{c})$ \ decreases as $%
\alpha $ increases in agreement with experimental evidences \cite{junod}. On
the other hand, in the range of temperatures close to $T_{c}$, and due to
thermally-created quasiparticle excitations, with roughly the same weight
for all directions of $\vec{k}$, qualitative differences with respect to
c-BCS behavior are not found and the specific heat jump has a sharp maximum
at $T_{c}$. We get a linear dependence of $\Delta C$ on $\Delta (0)$ instead
of the quadratic dependence reported in \cite{ausloos}. In this region a
broadened step-like discontinuity, as it has been found in some CS, is
presented. For UR $R_{2}=\Delta C(T_{c})/C_{en}(T_{c})$ \ we found that even
though the value that we have obtained is higher than that of the c-BCS it
is not enough to be in agreement with experimental reported that ranges
between 2 and 4.

The reduced thermodynamic critical magnetic field $H_{c}(T)/H_{c}(0)$ as a
function of $T/T_{c}$ shows a marked departure from the characteristic
parabolic dependence observed in CS. Another important feature of the curve $%
H_{c}(T)/H_{c}(0)$ is the change of curvature that takes place close to $%
0.3T_{c}$. The origin of this departure from the parabolic law is associated
with the temperature dependence of the specific heat in the superconducting
state. In ref.\cite{hao} a dependence like $\sqrt{1-(T/T_{c})^{2}}$ is
proposed instead of the parabolic law. Finally, a minimum value for the
ratio $R_{3}=H_{c}(0)/\sqrt{N_{0}/\Omega }\Delta (0)$ is obtained (3.83)
which is higher than that reports in the c-BCS theory.

Our two main conclusions are: First, the widely used logaritmic form of the
DOS in the van Hove scenario is a very disputable approximation. Second, the
van Hove scenario, even using the ``exact'' DOS although it enlightens quite
a lot, might need more elements to account properly for the thermodynamics
of HTS. It seems that other contributions neglected by the van Hove scenario
play an important role \cite{cucolo}.

\begin{center}
{\large {\bf Acknowledgments}}
\end{center}

D. Q. would like to thank the financial support and the hospitality of the
Department of Physics at CINVESTAV, Mexico, D.F.

\noindent *Permanent address: Departamento de F\'{\i}sica Te\'{o}rica,
Universidad de La Habana, 10400 La Habana, Cuba

\newpage

\begin{figure}[tbp]
\caption{Topology electronic band structure in the first Brillouin zone
following eq. (1) for $t^{\prime }=0$ (a), and $t^{\prime }=0.4t$ (b). The
energy is given in eV and the calculations correspond to $t=0.25$ eV.}
\label{fig:1}
\end{figure}

\begin{figure}[tbp]
\caption{Dependence of the critical temperature, $T_c$ (a) and the reduced
gap at zero temperature, $\Delta (0)$ (b) on the dimensionless interaction
constant $\protect\lambda =N_0V$, at $\protect\varepsilon _c=20$ meV and $%
\protect\alpha =0.01$ (solid line), 0.04 (dashed line) and 0.07 (dot-dashed
line).}
\label{fig:2}
\end{figure}

\begin{figure}[tbp]
\caption{Dependence of the ratio $R_1=2\Delta (0)/k_BT_c$ on the
dimensionless interaction constant $\protect\lambda =N_0V$, for the same
values of $\protect\alpha $ and $\protect\varepsilon _c$ of fig. 2. The
c-BCS and v-BCS universal ratios are presented by thin lines.}
\label{fig:3}
\end{figure}

\begin{figure}[tbp]
\caption{Critical temperature $T_c$ as a function on the dimensionless
interaction constant $\protect\lambda =N_0V$, for $\protect\varepsilon _c:20$
meV (curves 1), 35 meV (curves 2), and 50 meV (curves 3). The calculations
with the exact DOS, eq. (\ref{edos}) and its asymptotic behavior, eq. (\ref
{ados}) are plotted by solid and dot-dashed lines, respectively. The
approximate expression given by eq. (6) is represented in dashed line for $%
\protect\varepsilon _c=50$ meV. The value of $\protect\alpha =0$ has been
chosen for all calculated curves.}
\label{fig:4}
\end{figure}

\begin{figure}[tbp]
\caption{Dependence of the ratio $\Delta (T)/\Delta (0)$ on the reduced
temperature $T/T_c$. The $T\rightarrow 0$ (curve 1 by dashed lines) and $%
T\rightarrow T_c$ (curve 2 by dot-dashed lines) approaches according to eqs.
(\ref{gap0}) and (\ref{DTD0}), respectively are presented (for details see
text). For comparison the c-BCS result \protect\cite{muhlsch} is depicted in
solid circles. The calculation correspond to $\protect\alpha =0$.}
\label{fig:5}
\end{figure}

\begin{figure}[tbp]
\caption{Electronic specific heat in the normal state $C_{en}$ in units of $%
T $ as a function of $T$ for different values of $\protect\alpha $ (a); The
jump ratio $\Delta C(\protect\alpha )/C_{en}(T_c)$ as a function of $\protect%
\alpha $ for different values of $\protect\varepsilon _c$ at $\protect%
\lambda =0.1$ (b). For comparison the c-BCS and the v-BCS universal ratios
are presented by thin lines.}
\label{fig:6}
\end{figure}

\begin{figure}[tbp]
\caption{Temperature dependence of $H_c(T)/H_c(0)$ on $T/T_c$ obtained from
eq. (\ref{hc}) with the DOS given by eq. (\ref{edos}) for $\protect%
\varepsilon _c=$35 meV, $\protect\lambda =$0.1 (curve 1); $\protect%
\varepsilon _c=$20 meV, $\protect\lambda =$0.1 (curve 2); $\protect%
\varepsilon _c=$35 meV, $\protect\lambda =$0.02 (curve 3). For comparison
the empirical parabolic law (curve 4) of the c-BCS (see ref. \protect\cite
{schri}) is shown.}
\label{fig:7}
\end{figure}

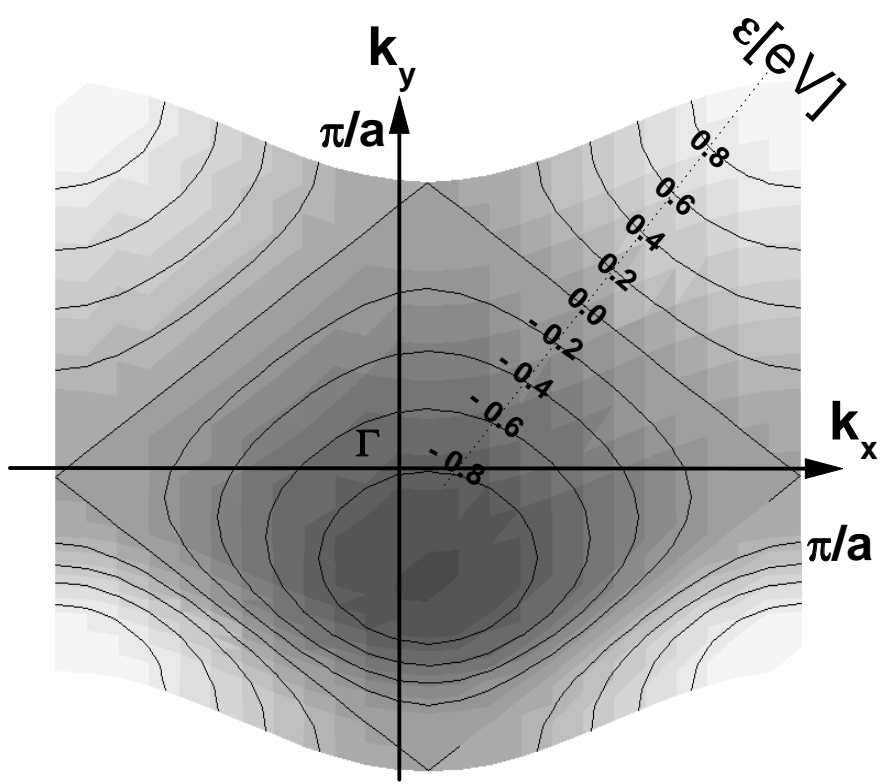
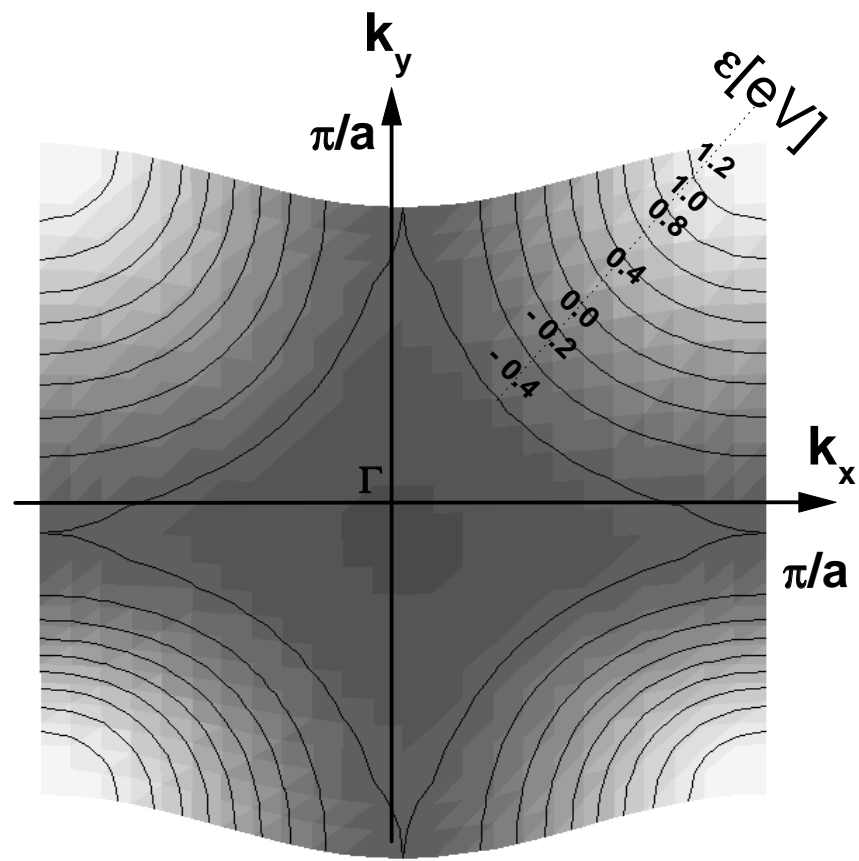
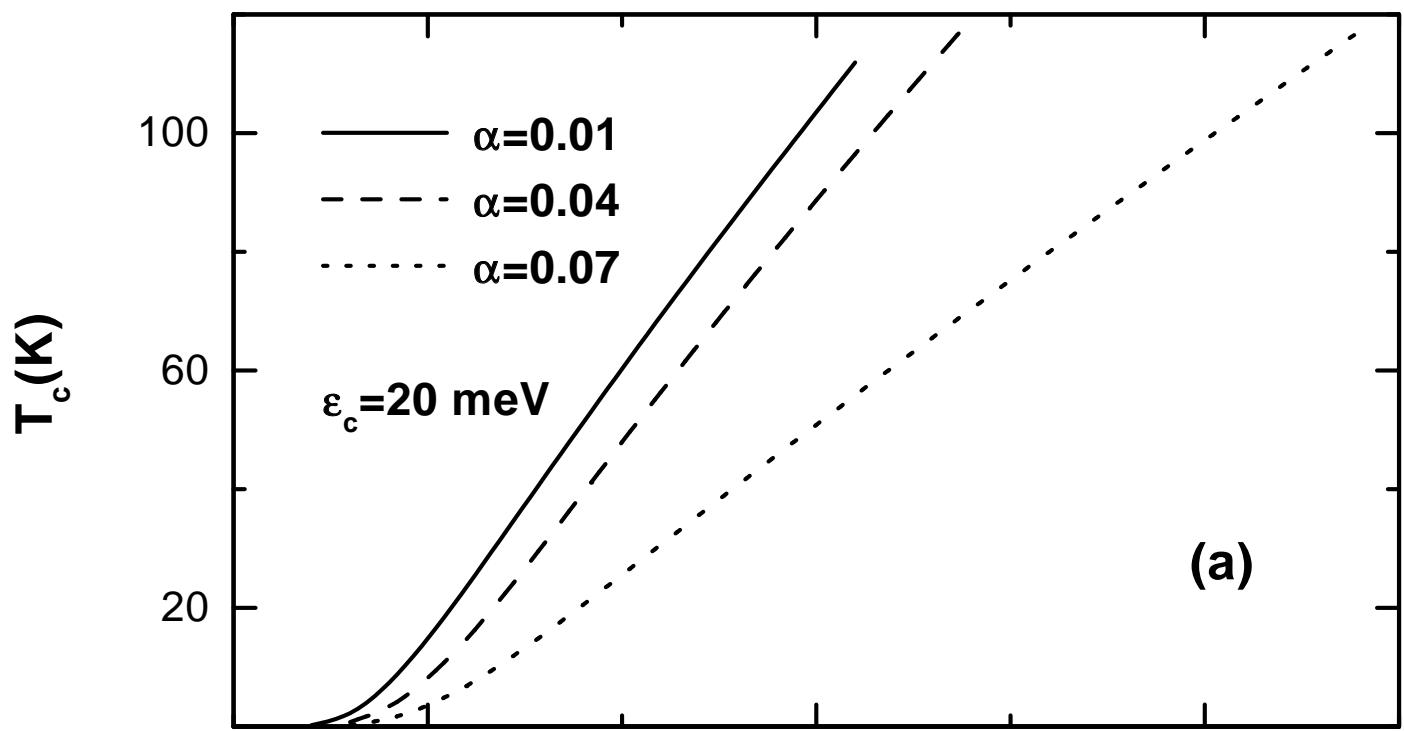
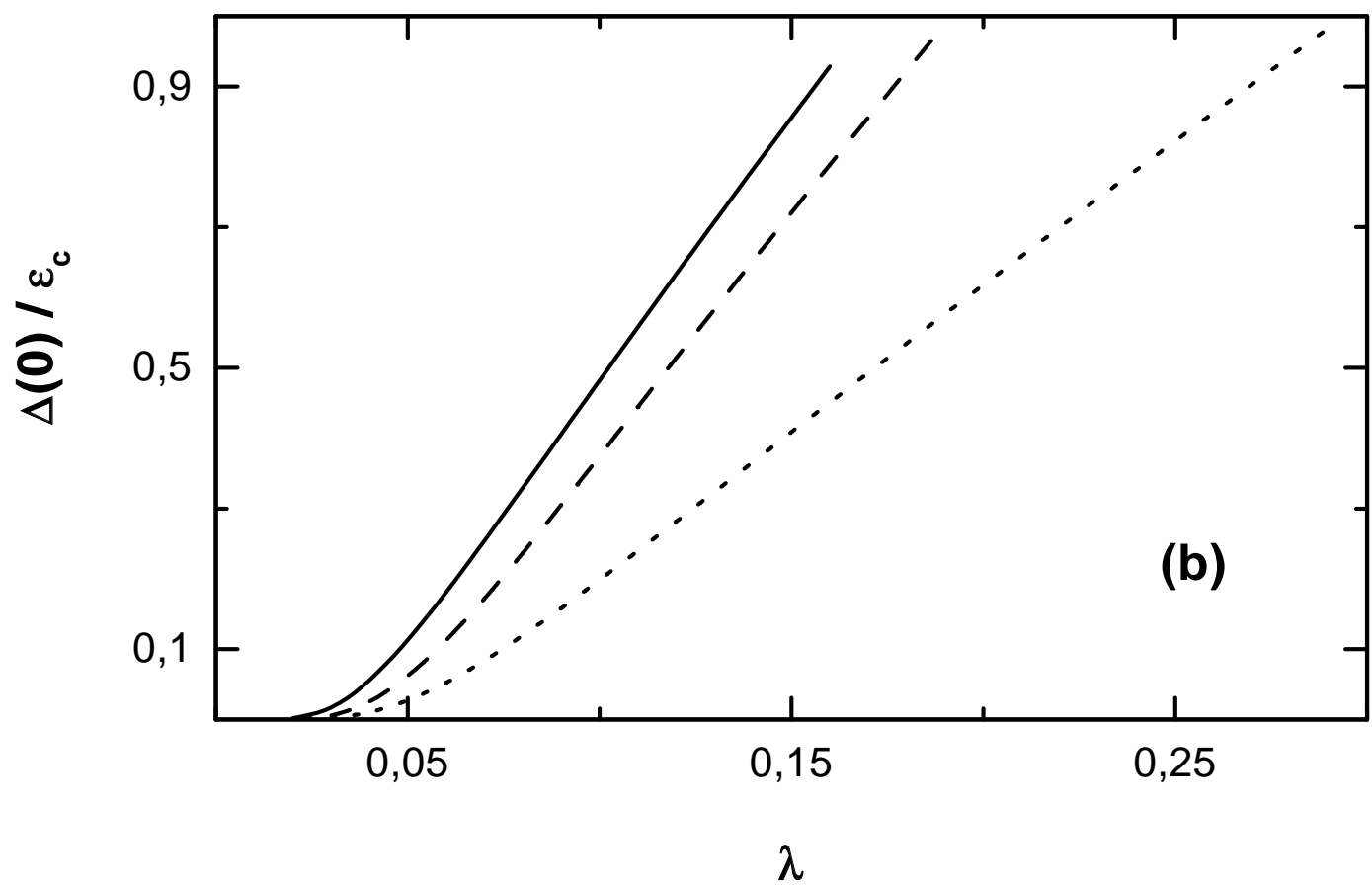
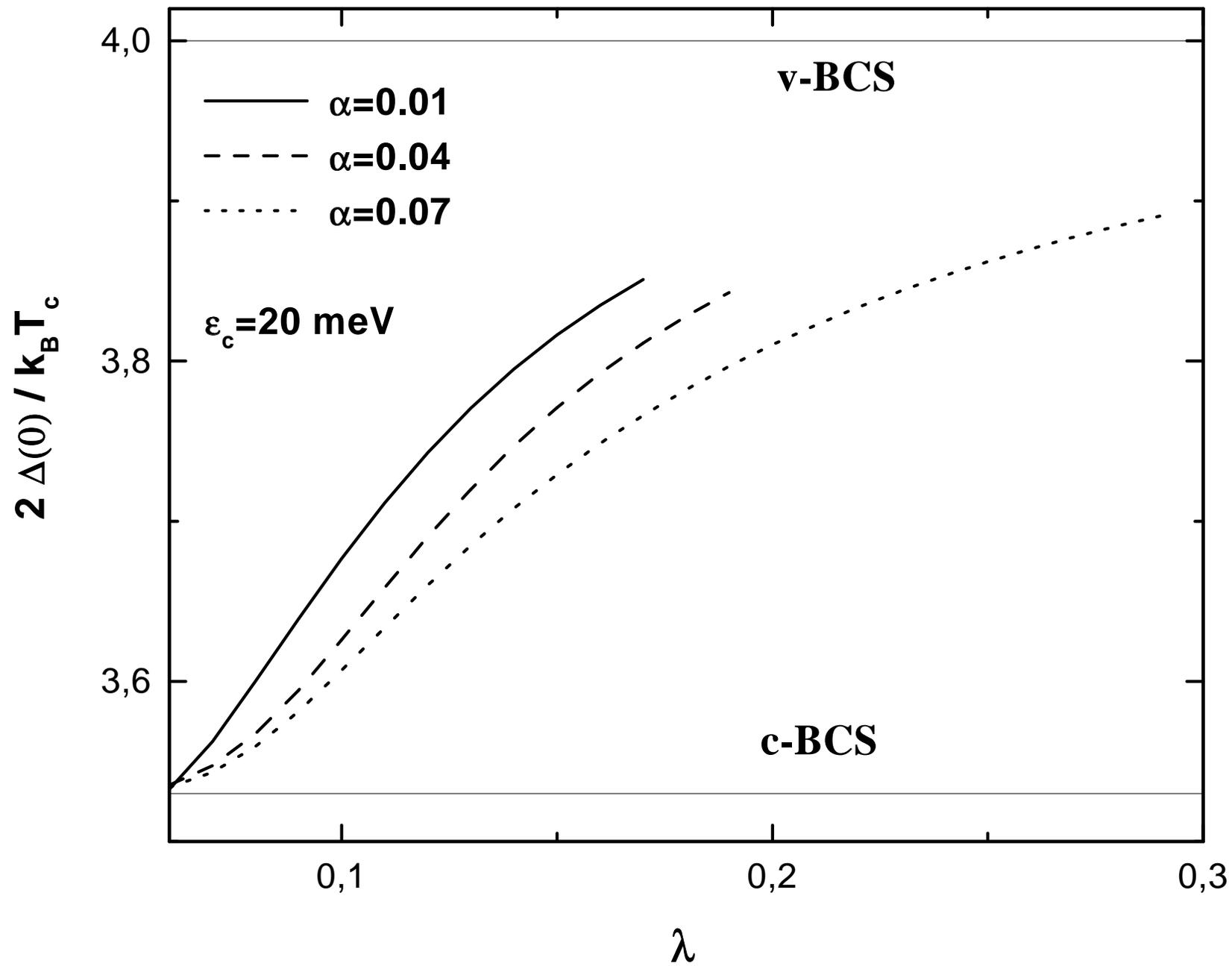
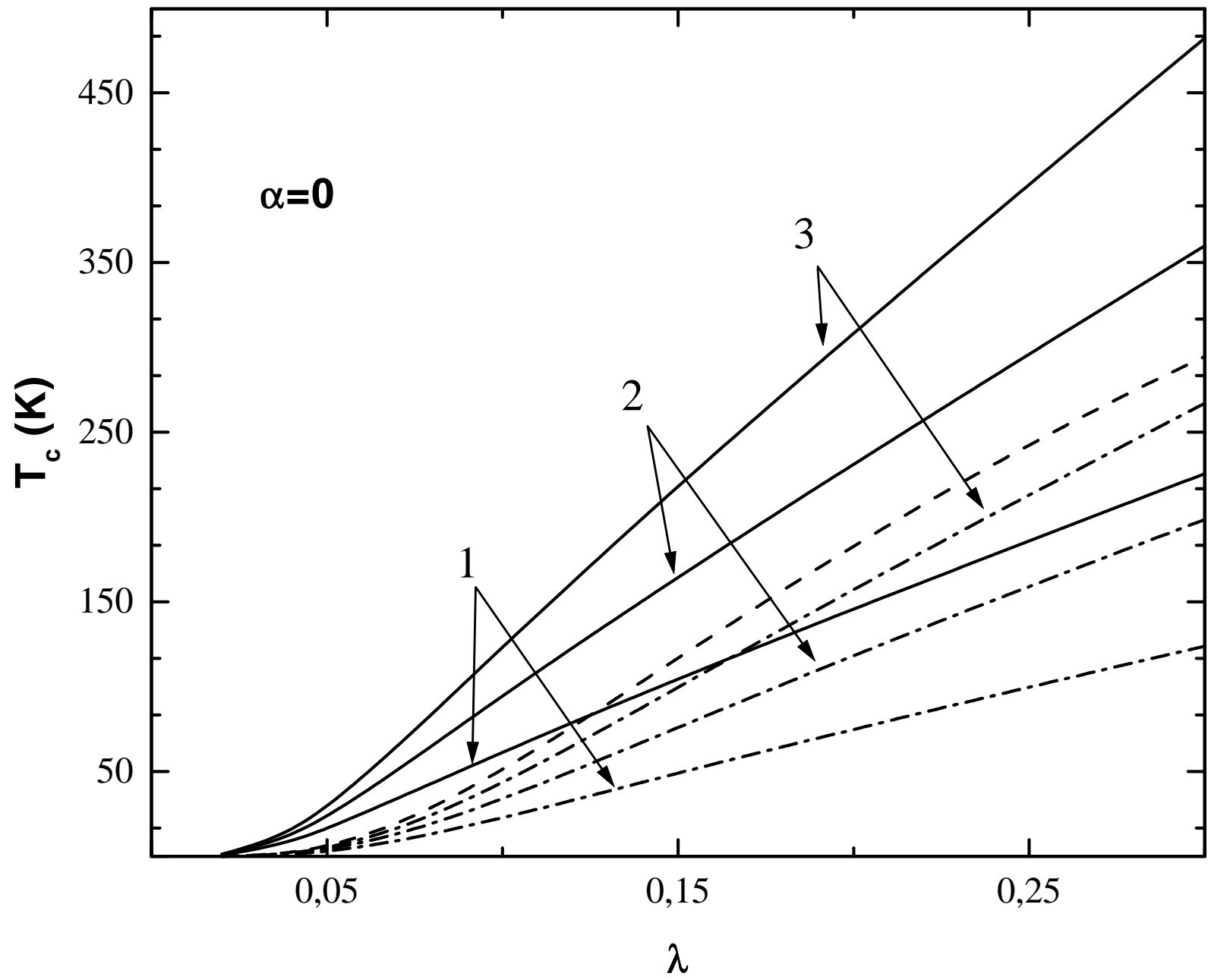
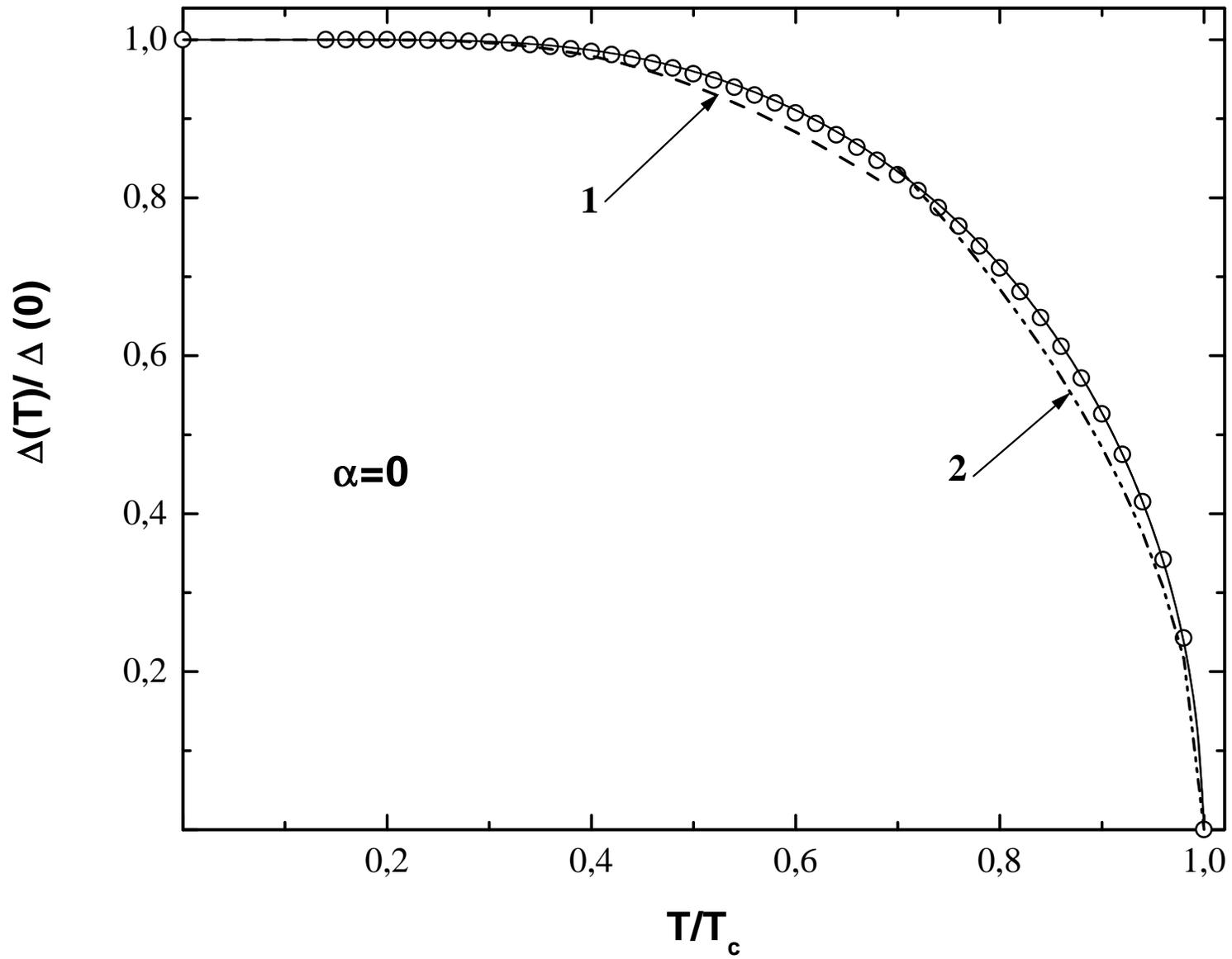
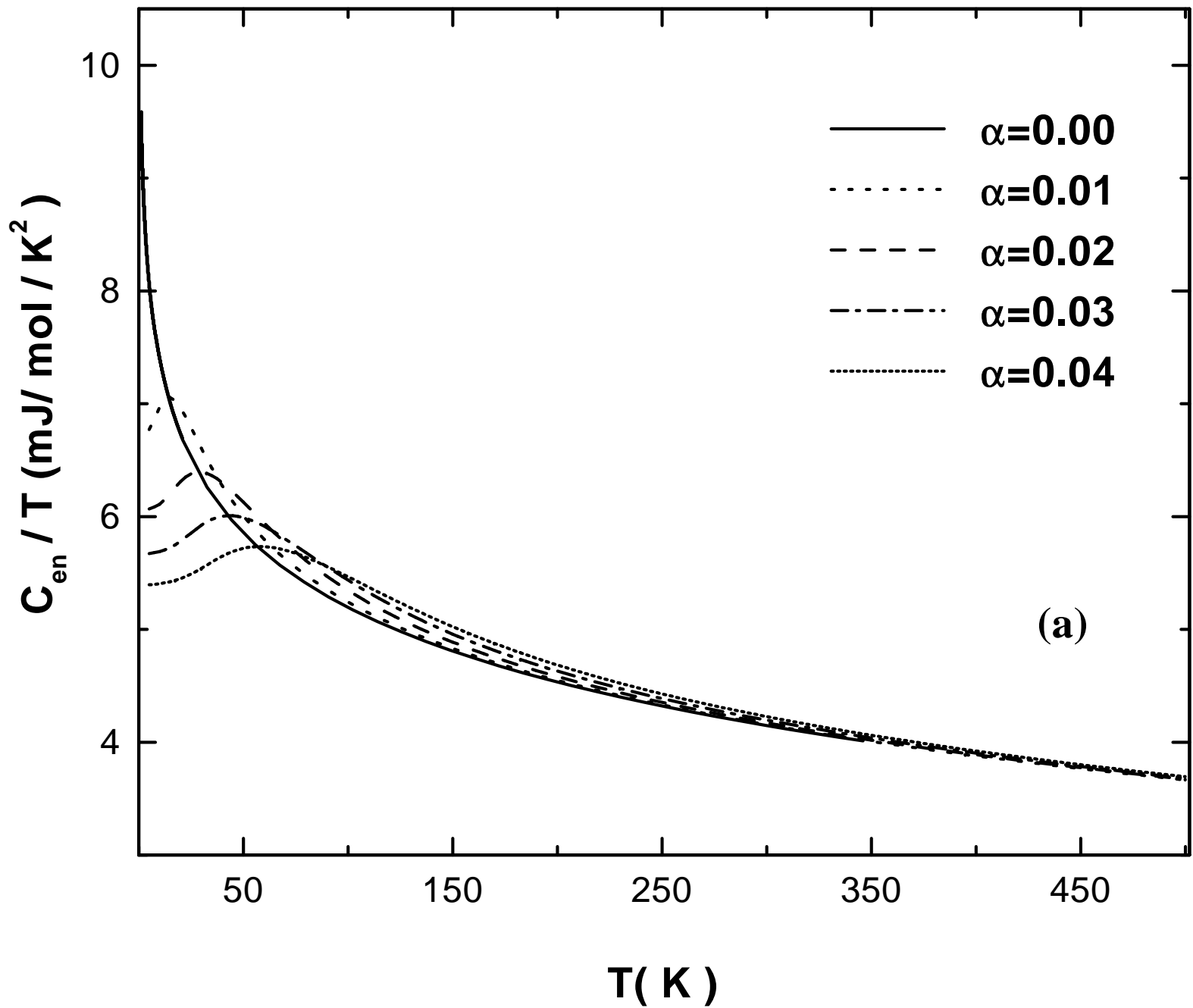
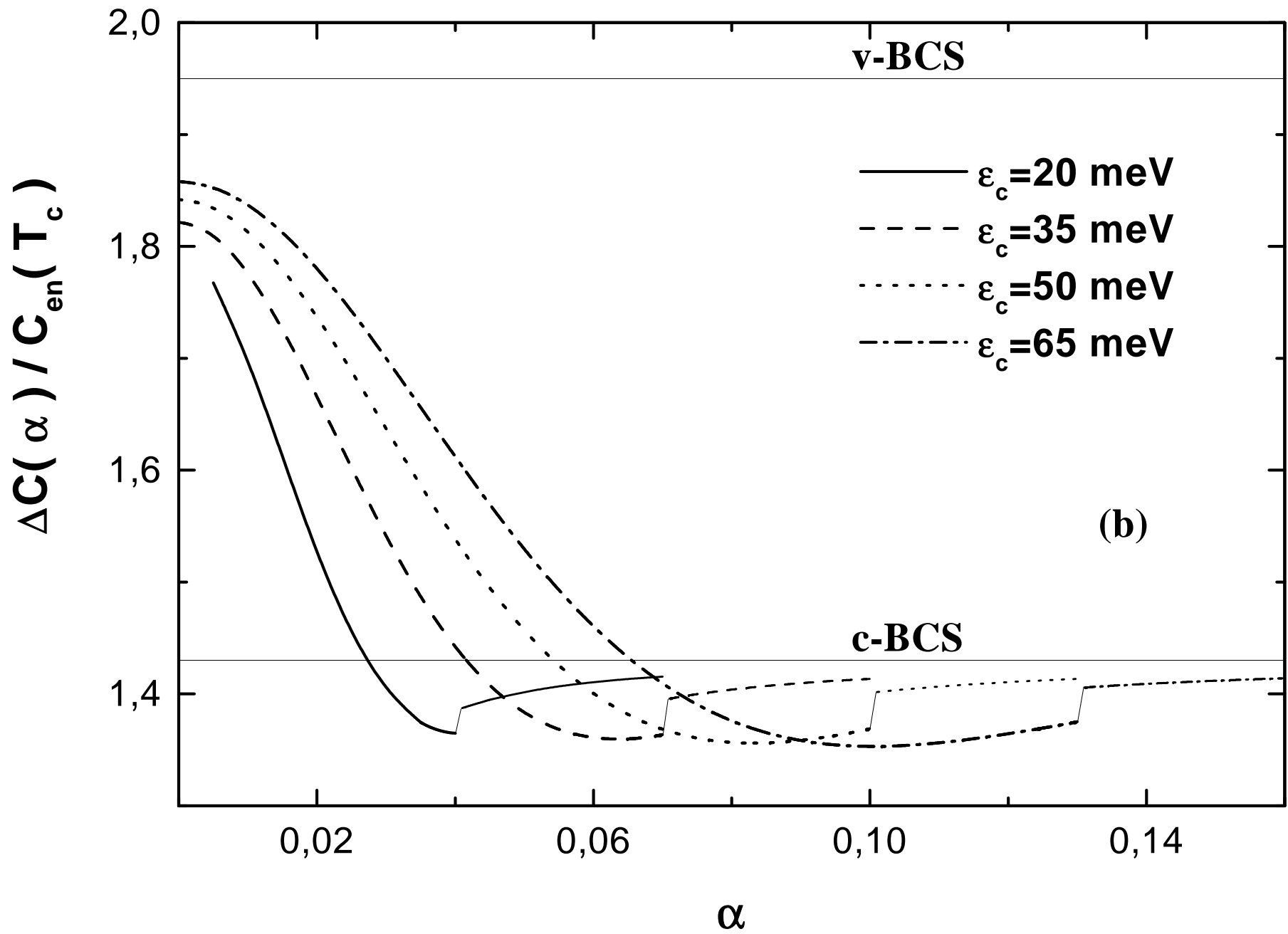
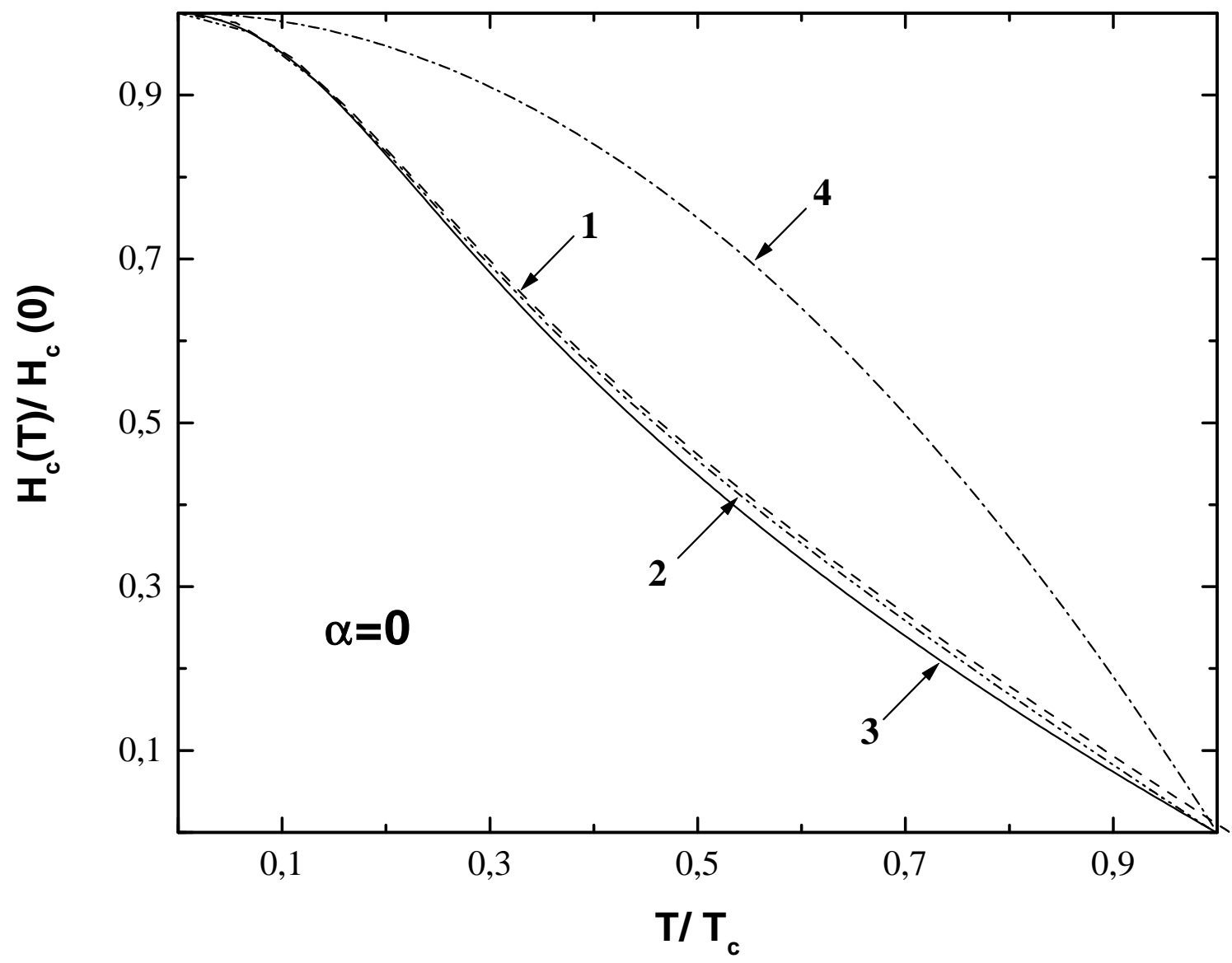
\end{document}